\documentclass[twocolumn,showpacs,preprintnumbers,amsmath,amssymb]{revtex4}
\usepackage{graphicx}% Include figure files
\usepackage{dcolumn}% Align table columns on decimal point
\usepackage{bm}% bold math
\usepackage{amsmath}
\usepackage{amsfonts}
\begin{document}

\preprint{APS/123-QED}

\title{Degeneracy Breaking of Hydrogen Atom}% Force line breaks with \\

\author{Agung Trisetyarso}
 \email{trisetyarso07@a8.keio.jp}
\affiliation{Department of Applied Physics and Physico-Informatics, Keio University, Yagami Campus\\
3-14-1 Hiyoshi, Kohoku-ku, Yokohama-shi, Kanagawa-ken 223-8522, Japan} 
 %\altaffiliation[Also at ]{Applied Physics and Physico-Informatics, Keio University.}%Lines break automatically or can be forced with \\

\author{Pantur Silaban}
 \email{psilaban@fi.itb.ac.id}
\affiliation{Department of Physics, Bandung Institute of Technology Bandung 40132, Indonesia}%

\date{\today}% It is always \today, today,

\begin{abstract}
The three dimensional rotation group, SO(3), is a symmetry group of the normal hydrogen atom. Each reducible representation of this group can be associated with a degenerate energy level. If this atom is placed in an external magnetic field, the interaction between the orbital magnetic moment with this field will lead to a symmetry breaking where the symmetry group of the atom is a new group distinct from the SO(3) group. This phenomenon describes the normal Zeeman effect, where a degenerate energy level splits into several new energy levels. It is explicitly shown that each of the new energy levels can be associated with an irreducible representation of the new symmetry group. 

%Valid PACS numbers may be entered using the \verb+\pacs{#1}+ command.

\end{abstract}

\pacs{Valid PACS appear here}% PACS, the Physics and Astronomy
\maketitle

\section[INTRODUCTION]{\label{sec:level1}INTRODUCTION}

Generally, every system such as an atom or molecule, which has uniquely symmetrical behavior, can always be represented in a degenerate state; i.e., energy level of the system represents an eigenvalue of Hamiltonian operator for the system which has a set of eigenvector. The simplest example is a hydrogen atom.

According to Bohr model, hydrogen atom has a spherical symmetry property, i.e., Hamiltonian operator for the atom is invariant over three-dimensional group transformation, SO(3). This symmetrical property causes for every principal quantum number, $n$=1,2,..., there is only one energy level for every orbital quantum number, $l$= 1, ..., $n$-1, and and for every magnetic quantum number, $m$= $-l$, $-l+1$, ... ,$l$. The energy level is given by:
\begin{equation}
E_{n,l,m}=-\frac{13,6}{n^{2}}\textbf{eV}
\end{equation}

\noindent
which is independent on $l$ and $m$.

If this hydrogen atom is placed in an external magnetic field, the spherical symmetry will be broken. This symmetry breaking causes hyperfine structure of energy level $E_{n,l,m}$ to the new several energy level. If one consider the interaction between orbital magnetic moment with the external magnetic field, so every energy level of $E_{n,l,m}$ will split into (2$l$+1) of the new energy level, which are represented by:
\begin{equation}
E_{n,l,m}=-l, E_{n,l,m}=-l+1, ... , E_{n,l,m}=l 
\end{equation}  

This paper proposes to formulate how hyperfine splitting of Hydrogen atom energy level placed in external magnetic field, in the form of three-dimensional rotation group representation.

Section II will explain about basic concept of group theory, especially the meaning of reducible representation and irreducible representation of continuous group or Lie group, which is useful to formulate symmetrical behavior of a system. Decomposition of a reducible representation into several irreducible representations is seemed as symmetry breaking. For a hydrogen atom, decomposition of reducible representation represented in three-dimensional rotation group into number of several irreducible representations will be useful to explain splitting of degenerate level into several new energy levels.

Section III clarifies Schr\"odinger and Heissenberg pictures associated to normal Zeeman effect. Schr\"odinger picture shows state vector dynamics and Heissenberg picture describes physical quantity associated to the hydrogen atom. In the case of hydrogen atom placed in an external magnetic field, the state vector is state vector of angular momentum operator and the physical quantity is angular momentum which is represented by generators of a three-dimensional rotation group.

Section IV builds structure of symmetry group for a hydrogen atom placed in an external magnetic field which is parallel to $z$ axis. This symmetry group is shown as direct product from one-dimensional unitary group, $U(1)$, with discrete group, ${\mathcal G}=\{E,I\}$, where $E$ represents identity operator and $I$ express inversion operator on $xy$-plane.

Section V provides conclusion emphazising that group structure built for normal Zeeman effect can be developed to explain anomalous Zeeman effect. Normal Zeeman effect is resulted by interaction between orbital magnetic moment from electron under the influence of external magnetic field, while anomalous Zeeman effect is affected by the total magnetic moment, that is orbital magnetic moment added with spin magnetic moment, and external magnetic field. This kind of interaction causes hyperfine splitting of hydrogen atom spectrum.   

\section[REPRESENTATION OF A GROUP SYMMETRY]{\label{sec:level2}REPRESENTATION OF A GROUP SYMMETRY}

Suppose $V$ is a $N$-dimensional linear vector space with basis vector $f_{\alpha}$,$\alpha=$1,2,...,$N$. Consider a group ${\mathcal G}$ of linear transformation $T_{k}$, $k$=1,2,..., so that

\begin{equation}
T_{k}f_{\alpha}=D_{\beta\alpha}(T_{k})f_{\beta}
\end{equation} 

Consequently, a set of ${\mathcal M}$ or ${\mathcal M}$=$\{D_{\beta\alpha}(T_{k})f_{\beta}, k=1,2,...\}$ of $N \times N$-dimensional matrices is resulted. Set of ${\mathcal M}$ defines a representation of  group ${\mathcal G}$ that is, if
\begin{equation}
T_{k}=T_{l}T_{m}
\end{equation}
only if
\begin{equation}
D_{\beta\alpha}=D_{\beta\gamma}(T_{l})D_{\gamma\alpha}(T_{m})
\end{equation}
$V$ vector space is called a \textit{representation space}, and vector $f_{\alpha}$, $\alpha$=1,2,...,$N$, forms basis system for representation $D_{\beta\alpha}(T_{k})$. Representation $D_{\beta\alpha}(T_{k})$ is named $N$-dimensional representation, and usually called that \textit{$f_{\alpha}$ vector transformation is under representation $D_{\beta\alpha}(T_{k})$ of group ${\mathcal G}$}. 

Furthermore, $D_{\beta\alpha}(T_{k})$ representation will be written as $D(T)$. Two representation, $D_{1}(T)$ and $D_{2}(T)$ is called \textit{equivalent} if there is a nonsingular matrix ${\mathcal P}$ fulfill
\begin{equation}
D_{1}(T)={\mathcal P}^{-1}D_{2}(T){\mathcal P}
\end{equation}    

Suppose $N$-dimensional representation, $D(T)$, is equivalent to matrix $X(T)$, where
\begin{equation}
X(T)=\left( \begin{array}{cccc}
X_{1}(T) & 0 & 0 & 0 \\
0 & X_{2}(T) & 0 & 0 \\
0 & 0 & ... & ... \\
0 & 0 & ... & X_{k}(T) \end{array} \right) 
\end{equation} 

Subsequently, representation $D(T)$ is called an \textit{irreducible representation}. In this case, representation $D(T)$ can be written as
\begin{equation}
D(T)=X_{1}(T) \oplus X_{2}(T) \oplus ... \oplus X_{k}(T)
\end{equation} 

If the dimension of matrices $X_{1}(T), X_{2}(T), ... , X_{k}(T)$ respectively are $N_{1}, N_{2}, ... , N_{k}$ so that 
\begin{equation}
N=N_{1} \oplus N_{2}\oplus ... \oplus N_{k}
\end{equation} 

\textit{Representation D(T) is called an irreducible representation if D(T) is not a reducible representation}.

$N$-dimensional Vector space $V$ is called \textit{reducible vector space} if its basis vector, $f_{\alpha}$,$\alpha$=1,2,...,$N$, transformation is under a reducible representation. If its vector basis transformation is under an irreducible representation, consequently it is called \textit{irreducible vector space}.

The importance of representation in physics, especially in quantum mechanics, can be seen in several theorems as provided below.

$\textbf{Theorem 1}$ {\it Suppose Hamilton operator ${\mathcal H}$ of a system is invariant respect to a transformation group ${\mathcal G}$. Let for every eigen vector of ${\mathcal H}$ associated with one eigenvalue so that the eigenfunctions are defined by a basis for representation of group ${\mathcal G}$}.

$\textbf{Theorem 2}$ {\it Suppose Hamilton operator ${\mathcal H}$ of a system is invariant respect to transformation group ${\mathcal G}$, so that every eigenvectors transforming under one irreducible representation of ${\mathcal G}$ will be associated with the equivalent energy level.}

In Zeeman effect analysis, we only need \textit{compact group}. Every representation of compact group are equivalent with unitary representation.

A \textit{compact group} is a transformation group which is continuous with finite parameters; for example, a $N$-dimensional rotation group is a compact group due to the parameter is given by rotation angle between 0 and 2$\pi$.

A \textit{non-compact group} is a continuous transformation group with infinite parameter; for example, a $N$-dimensional translation group is a non-compact group because its parameter is given by the vector translation which is between 0 and $\infty$.

Representation of compact group and representation of non-compact group have different characteristics which is necessary to be understood from topology consideration, which is not mentioned in this paper. 

\section[SCHR\"ODINGER AND HEISSENBERG PICTURES OF NORMAL ZEEMAN EFFECT]{\label{sec:level3}SCHR\"ODINGER AND HEISSENBERG PICTURES OF NORMAL ZEEMAN EFFECT}

Hamilton operator ${\mathcal H}$ for a hydrogen atom which is placed in an external magnetic field $\stackrel{\rightarrow}{B}$ is given by

\begin{equation}
{\mathcal H}={\mathcal H}_{0}-\frac{e}{2mc}\stackrel{\rightarrow}{B}.\stackrel{\rightarrow}{L}
\end{equation}
\noindent
where

\begin{equation}
{\mathcal H}_{0}=-\frac{\hbar}{2m}\nabla^{2}+e\phi
\end{equation}

with $\stackrel{\rightarrow}{L}$ represents angular momentum operator, $m$ is the mass of electron, $e$ is the charge of electron, $e\phi$ is electrostatic potential energy of electron, and $\mu_{0}$ = $\frac{e\hbar}{2mc}$=-0,9273x10$^{-20}$$\frac{erg}{gauss}$ is electron orbital magnetic moment.  

By choosing $\stackrel{\rightarrow}{B}$ in the direction of $z$ axis, the Hamilton operator will be in the form

\begin{equation}
{\mathcal H}={\mathcal H}_{0}-\mu_{0}B_{z}L_{z}
\end{equation}
\noindent
where $L_{z}$ is an angular momentum operator in $z$ axis direction.

Perturbation potential, -$\mu_{0}B_{z}L_{z}$, will cause hyperfine splitting of energy level of this hydrogen atom. This effect is called normal Zeeman effect.

In order to explain Schr\"odinger and Heissenberg pictures which will be associated with the normal Zeeman effect, we need eigenvector $|\psi\rangle$ of angular momentum operator, $\stackrel{\rightarrow}{L}$, which is expressed in polar coordinate ($\theta,\phi$). In polar coordinate ($\theta,\phi$), angular momentum operator, $\stackrel{\rightarrow}{L}$, is given by:
\begin{equation}
\stackrel{\rightarrow}{L}=L_{x}~sin\theta~cos\phi+L_{y}~sin\theta~sin\phi~+L_{z}~cos\theta
\end{equation} 
\noindent
where

\begin{equation}
L_{x}=i\hbar\left( \begin{array}{ccc}
0 & 0 & 0 \\
0 & 0 & -1 \\
0 & 1 & 0 \end{array} \right), L_{y}=i\hbar\left( \begin{array}{ccc}
0 & 0 & 1 \\
0 & 0 & 0 \\
-1 & 0 & 0 \end{array} \right)
\end{equation}
\begin{equation} 
L_{z}=i\hbar\left( \begin{array}{ccc}
0 & -1 & 0 \\
1 & 0 & 0 \\
0 & 0 & 0 \end{array} \right)  
\end{equation}

Consequently
\begin{equation}
\stackrel{\rightarrow}{L}=i\hbar\left( \begin{array}{ccc}
0 & -cos\theta & sin\theta~sin\phi \\
cos\theta & 0 & -sin\theta~cos\phi \\
-sin\theta~sin\phi & sin\theta~cos\phi & 0 \end{array} \right)
\end{equation}
\noindent
By solving eigenvalue problem
\begin{equation}
\stackrel{\rightarrow}{L}|\psi\rangle=\lambda|\psi\rangle
\end{equation}

\noindent
resulting ${\mathcal \lambda}=\pm \hbar$. For ${\mathcal \lambda}=+ \hbar$, we can obtain a normalized eigen vector
\begin{equation}
|\psi\rangle=|\psi_{+}\rangle=\frac{1}{\sqrt{2}}\left( \begin{array}{c}
-cos\theta~cos\phi+isin\phi\\
-cos\theta~sin\phi-icos\phi\\
sin\theta\end{array} \right) 
\end{equation}
\noindent
Similarly, for ${\mathcal \lambda}=- \hbar$
\begin{equation}
|\psi\rangle=|\psi_{-}\rangle=\frac{1}{\sqrt{2}}\left( \begin{array}{c}
-cos\theta~cos\phi-isin\phi\\
-cos\theta~sin\phi+icos\phi\\
sin\theta\end{array} \right) 
\end{equation}
\noindent 
Schr\"odinger picture formulates dynamics of $|\psi\rangle$, that is the change of $|\psi\rangle$ respect to time $t$, which is given by the equation

\begin{equation}
i\frac{\partial}{\partial t}|\psi\rangle=-\frac{e}{2mc}B_{z}L_{z}|\psi\rangle
\end{equation}   
\noindent
By the use of $|\psi_{+}\rangle$, one can obtain
\begin{eqnarray*}
\frac{\partial}{\partial t}\left( \begin{array}{c}
-cos\theta~cos\phi+isin\phi\\
-cos\theta~sin\phi-icos\phi\\
sin\theta\end{array} \right)~\\ 
=\frac{-eB_{z}}{2mc}\left( \begin{array}{c}
cos\theta~sin\phi+icos\phi\\
-cos\theta~cos\phi+isin\phi\\
0\end{array} \right) 
\end{eqnarray*}

\noindent
which gives the equations below
\begin{widetext}
\begin{eqnarray}
cos\theta~sin\phi~\frac{d\phi}{dt}+sin\theta~cos\phi~\frac{d\theta}{dt}+icos\phi~\frac{d\phi}{dt}=\frac{-eB_{z}}{2mc}(cos\theta~sin\phi+icos\phi)\\
-cos\theta~cos\phi~\frac{d\phi}{dt}+sin\theta~sin\phi~\frac{d\theta}{dt}+isin\phi~\frac{d\phi}{dt}=\frac{-eB_{z}}{2mc}(-cos\theta~cos\phi+isin\phi)\\
cos\theta~\frac{d\theta}{dt}=0
\end{eqnarray}
\end{widetext}
\noindent
The above equations results
\begin{eqnarray}
\theta=constant\\
\phi(t)=\phi(0)-(\frac{eB_{z}}{2mc})t
\end{eqnarray} 
\noindent
where $\phi$(0) is $\phi$ angle on time $t$=0. This result shows that state vector rotates respect to $z$ axis with angular velocity $\frac{-eB_{z}}{2mc}$. The same conclusion can be obtained by the use of $|\psi_{-}\rangle$.

Heissenberg picture formulates the dynamics of angular momentum operator $L_{x}$, $L_{y}$, and $L_{z}$, that is the change of those operators respect to time $t$, which are represented by the equations
\begin{eqnarray*}
i\frac{dL_{x}}{dt}=[L_{x},{\mathcal H}']\\
i\frac{dL_{y}}{dt}=[L_{y},{\mathcal H}']\\
i\frac{dL_{z}}{dt}=[L_{z},{\mathcal H}']
\end{eqnarray*}

\noindent
where
\begin{equation*}
{\mathcal H}'=-\frac{eB_{z}\hbar}{2mc}L_{z}
\end{equation*}
\noindent
Due to the relation
\begin{equation}
[L_{x},L_{y}]=i\hbar L_{z}, [L_{y},L_{z}]=i\hbar L_{x}, [L_{z},L_{x}]=i\hbar L_{y} 
\end{equation}  
\noindent 
subsequently, the equations result
\begin{eqnarray}
\frac{dL_{x}}{dt}=\frac{e\hbar B_{z}}{2mc}L_{y}\\
\frac{dL_{y}}{dt}=-\frac{e\hbar B_{z}}{2mc}L_{x}\\
\frac{dL_{z}}{dt}=0
\end{eqnarray}
\noindent
so that
\begin{eqnarray}
L_{z}=~constant\\
\frac{d^{2}L_{x}}{dt^{2}}=-\frac{e^{2}\hbar^{2}B_{z}^{2}}{4m^{2}c^{2}}L_{x}\\
\frac{d^{2}L_{y}}{dt^{2}}=-\frac{e^{2}\hbar^{2}B_{z}^{2}}{4m^{2}c^{2}}L_{y}
\end{eqnarray}
\noindent
which have the solutions
\begin{eqnarray}
L_{x}(t)=Ae^{i\omega t}+Be^{-i\omega t}\\
L_{y}(t)=Ce^{i\omega t}+De^{-i\omega t}
\end{eqnarray}
\noindent
where
\begin{equation*}
\omega~=~\frac{e\hbar B_{z}}{2mc} 
\end{equation*}
\noindent
and $\it{A, B, C}$ and $\it{D}$ respectively are a constant. It is clear that angular momentum in $z$ axis direction is also a constant, while angular momentum in the direction of $x$ axis and $y$ axis respectively has precision with velocity magnitude $\frac{e\hbar B_{z}}{2mc}$.

\section[SYMMETRY GROUP OF NORMAL ZEEMAN EFFECT]{\label{sec:level4}SYMMETRY GROUP OF NORMAL ZEEMAN EFFECT}

In order to explain hyperfine splitting in normal Zeeman effect, we will use the well-known theorem in group theory and also its application is widely used in quantum mechanics.

$\textbf{Theorem~3}$ \textit{Suppose Hamilton operator ${\mathcal H}$ of a system is given by}
\begin{equation}
{\mathcal H}={\mathcal H}_{0}+{\mathcal H}'
\end{equation}
\noindent
\textit{where ${\mathcal H}'$ represents a perturbation operator. Let ${\mathcal H}, {\mathcal H}_{0}$ and  ${\mathcal H}'$ are invariant respect to transformation group ${\mathcal G}$ which its generators are expressed by $X_{\alpha},\alpha$=1,2,..., that is}
\begin{equation}
[{\mathcal H},X_{\alpha}]=0,~[{\mathcal H}_{0},X_{\alpha}]=0,~[{\mathcal H}',X_{\alpha}]=0
\end{equation}
\noindent
\textit{If set of $|\psi_{k}\rangle$, $k$=1,2,...,$n$ are eigenvectors of ${\mathcal H}_{0}$ with one eigenvalue $\lambda$, that is}
\begin{equation}
{\mathcal H}_{0}|\psi_{k}\rangle=\lambda|\psi_{k}\rangle
\end{equation}
\noindent
$\textit{and if $|\psi_{k}\rangle$ are transformed under representation}$
\begin{equation}
D=D^{(1)}\oplus D^{(2)}\oplus D^{(3)}\oplus ... D^{(k)}
\end{equation}
\noindent
\textit{where $D^{(k)}$ represents an irreducible component of representation $D$, subsequently the number of new energy levels which is resulted from energy hyperfince splitting which is represented by eigenvalue $\lambda$ is $n$. Every $D^{(k)}$ is associated with one new energy level.}

This theorem helps us to understand hyperfine splitting of a degeneracy energy level to several new energy levels. Hamilton operator
\begin{equation}
{\mathcal H}_{0}=-\frac{\hbar^{2}}{2m}\nabla^{2}-\frac{e^{2}}{r}
\end{equation}   
\noindent
of a hydrogen atom is invariant respect to rotation group $SO(3)$.

Because the perturbation operator
\begin{equation}
{\mathcal H}'=-\frac{e\hbar}{2mc}B_{z}L_{z}
\end{equation} 
\noindent
is only invariant respect to rotation in $z$ axis direction, that is
\begin{equation}
[{\mathcal H}',L_{x}]\neq0,~[{\mathcal H}',L_{y}]\neq0,~[{\mathcal H}',L_{z}]=0
\end{equation} 
\noindent
consequently, new symmetry group of the system is \textit{direct product group} which is given by $U(1)\otimes\{E,I\}$, where $U(1)$ is a one-dimensional unitary group representing rotation in $z$ axis direction, $E$ is identity operator and $I$ is inversion operator which is defined by
\begin{equation}
I:(x,y,z)\rightarrow(-x,-y,-z)
\end{equation}
To clarify energy hyperfine splitting, let us observe again a representation of three-dimensional rotation group. Due to the algebra structure of this group is given by:
\begin{equation}
[L_{a},L_{b}]=i\epsilon_{abc}L_{c}
\end{equation}
\noindent
where $a,b,c$=1,2,3, and $\epsilon_{abc}$ represents a three-dimensional permutation symbol, that is
\begin{equation}
\epsilon_{abc}=\left\{ 
\begin{array}{cc}
+1~if~abc~is~even~permutation\\
0~if~two~or~more~of~a,~b,~c~are~same \\
-1~if~abc~is~odd~permutation
\end{array}
\right\}\end{equation}
\noindent
Generally, representation of three-dimensional rotation group is described by following theorem

$\textbf{Theorem~4}$ \textit{Suppose $D[g(\stackrel{\rightarrow}{\alpha})]$ is a matrix associated with an element $g(\stackrel{\rightarrow}{\alpha})$ of a three-dimensional rotation group where $\stackrel{\rightarrow}{\alpha}$=($\alpha_{x}$, $\alpha_{y}$, $\alpha_{z}$) is a parameter which is very small or $\stackrel{\rightarrow}{\alpha}$$\approx$ 0. Consequently, set of all matrices
\begin{equation}
D[g(\stackrel{\rightarrow}{\alpha})]=e^{i\stackrel{\rightarrow}{L}.\stackrel{\rightarrow}{\alpha}}
\end{equation}
defines a representation of the group
}.  
\\
\\
\textit{Proof}

Consider $D[g(\stackrel{\rightarrow}{\alpha})]=e^{i\stackrel{\rightarrow}{L}.\stackrel{\rightarrow}{\alpha}}$ and $D[g(\stackrel{\rightarrow}{\beta})]=e^{i\stackrel{\rightarrow}{L}.\stackrel{\rightarrow}{\beta}}$, where $\stackrel{\rightarrow}{\alpha}$ and $\stackrel{\rightarrow}{\beta}$ express very small parameters. If
\begin{equation}
g(\stackrel{\rightarrow}{\alpha})g(\stackrel{\rightarrow}{\beta})=g(\stackrel{\rightarrow}{\alpha}+\stackrel{\rightarrow}{\beta})
\end{equation}
\noindent
it can be shown that
\begin{equation}
D[g(\stackrel{\rightarrow}{\alpha})]D[g(\stackrel{\rightarrow}{\beta})]=D[g(\stackrel{\rightarrow}{\alpha}+\stackrel{\rightarrow}{\beta})]
\end{equation}
\noindent
Note that
\begin{eqnarray*}
D[g(\stackrel{\rightarrow}{\alpha})]D[g(\stackrel{\rightarrow}{\beta})]=e^{i\stackrel{\rightarrow}{L}.\stackrel{\rightarrow}{\alpha}}e^{i\stackrel{\rightarrow}{L}.\stackrel{\rightarrow}{\beta}}
\\=(1+i\stackrel{\rightarrow}{L}.\stackrel{\rightarrow}{\alpha})(1+i\stackrel{\rightarrow}{L}.\stackrel{\rightarrow}{\beta})
\\=1+i\stackrel{\rightarrow}{L}.(\stackrel{\rightarrow}{\alpha}+\stackrel{\rightarrow}{\beta})
\\=e^{i\stackrel{\rightarrow}{L}.(\stackrel{\rightarrow}{\alpha}+\stackrel{\rightarrow}{\beta})}
\end{eqnarray*}
\noindent
which means
\begin{equation*}
D[g(\stackrel{\rightarrow}{\alpha})]D[g(\stackrel{\rightarrow}{\beta})]=D[g(\stackrel{\rightarrow}{\alpha}+\stackrel{\rightarrow}{\beta})]
\end{equation*}
\noindent
representing that set of matrices $D[g(\stackrel{\rightarrow}{\alpha})]$ define a representation of the three-dimensional rotation group.

\noindent
Consider
\begin{eqnarray*}
L_{z}|\phi\rangle=\lambda|\phi\rangle\\
(L_{x}+iL_{y})|\phi\rangle=|\phi_{+}\rangle\\
(L_{x}-iL_{y})|\phi\rangle=|\phi_{-}\rangle
\end{eqnarray*}
subsequently $|\phi_{+}\rangle$ is an eigenvector of $L_{z}$ with eigenvalue ($\lambda$+1), and $|\phi_{-}\rangle$ is an eigenvector of $L_{z}$ with eigenvalue ($\lambda$-1). Suppose $l$ is maximum eigenvalue of $L_{z}$ with eigenvector $|\phi_{l}\rangle$, where $\langle\phi_{l}|\phi_{l}\rangle$=1.
\noindent
If $(L_{x}-iL_{y})|\phi_{l}\rangle=\mu_{l}|\phi_{l-1}\rangle$ with $\langle\phi_{l-1}|\phi_{l-1}\rangle$=1, one can obtain $L_{z}|\phi_{l-1}\rangle=(l-1)|\phi_{l-1}\rangle$. Furthermore, $(L_{x}-iL_{y})|\phi_{l-1}\rangle=\mu_{l-1}|\phi_{l-2}\rangle$, with $\langle\phi_{l-2}|\phi_{l-2}\rangle$=1, thus $L_{z}|\phi_{l-2}\rangle=(l-2)|\phi_{l-2}\rangle$ can be determined. Generally, it can be shown that $(L_{x}-iL_{y})|\phi_{l-k}\rangle=\mu_{l-k}|\phi_{l-k-1}\rangle$ with $\langle\phi_{l-k}|\phi_{l-k}\rangle$=1, consequently  $L_{z}|\phi_{l-k}\rangle=(l-k)|\phi_{l-k}\rangle$, where $k$=0,1,2, ... .    

Due to the finiteness of the number of $L_{z}$ different eigenvalue, the sequence of $|\phi_{l}\rangle$, $|\phi_{l-1}\rangle$, ... must be ended on certain eigenvector, for example $|\phi_{m}\rangle$, that is $(L_{x}-iL_{y})|\phi_{m}\rangle=0$, with  $\langle\phi_{m}|\phi_{m}\rangle$=1 and
\begin{equation}
\label{eq51}
L_{z}|\phi_{m}\rangle=(m)|\phi_{m}\rangle
\end{equation}    
\noindent
Because 
\begin{equation}
\label{eq52}
(L_{x}-iL_{y})|\phi_{m}\rangle=\mu_{m}|\phi_{m-1}\rangle
\end{equation}
means $\mu_{m}=$0. Furthermore, consider step-up operator $(L_{x}+iL_{y})|\phi_{l-1}\rangle=\frac{2\hbar l}{\mu_{l}}|\phi_{l}\rangle$. Thus, the vector of $(L_{x}+iL_{y})|\phi_{l-1}\rangle$ is equal to $|\phi_{l}\rangle$ vector, which means that there is a number $\alpha_{l}\textgreater$0 which suffices $(L_{x}+iL_{y})|\phi_{l-1}\rangle=\alpha_{l}|\phi_{l}\rangle$. Similarly, it can be shown that $(L_{x}+iL_{y})|\phi_{m}\rangle$ is a vector which is equivalent to $|\phi_{m+1}\rangle$ meaning that there is a number $\alpha_{m+1}\textgreater$0 which fulfills $(L_{x}+iL_{y})|\phi_{m}\rangle=\alpha_{m}|\phi_{m+1}\rangle$.    

It is clear that
\begin{equation}
(L_{x}+iL_{y})|\phi_{m}\rangle=\frac{2\hbar L_{z}}{\mu_{m+1}}|\phi_{m+1}\rangle+\frac{\alpha_{m+2}}{\mu_{m+1}}(L_{x}-iL_{y})|\phi_{m+2}\rangle
\end{equation}
\noindent
and by using equations (\ref{eq51}) and (\ref{eq52}), one can obtain
\begin{equation}
\label{eq54}
\frac{2(m+1)+(\mu_{m+2})(\alpha_{m+2})}{\mu_{m+1}}=\alpha_{m+1}
\end{equation}  

Because of $(L_{x}+iL_{y})|\phi_{l}\rangle$=0 thus $\alpha_{l+1}$=0 for $l$=$m$. Furthermore, because $\langle\phi_{m}|\phi_{m}\rangle$=$\langle\phi_{m-1}|\phi_{m-1}\rangle$=1 and $(L_{x}+iL_{y})^{\dagger}$ = $(L_{x}-iL_{y})$ where $(L_{x}+iL_{y})^{\dagger}$ represents hermitian conjugate of $(L_{x}+iL_{y})$, subsequently $\alpha_{m}\langle\phi_{m}|\phi_{m}\rangle$=$\mu_{m}\langle\phi_{m-1}|\phi_{m-1}\rangle$ which means $\alpha_{m}$=$\mu_{m}$.  

Equation (\ref{eq54}) is in the form
\begin{equation}
\mu_{m}^{2}-\mu_{m+1}^{2}=2m
\end{equation}
\noindent
By adding from $\alpha=m$ to $\alpha=+l$, that is
\begin{equation*}
\sum_{\alpha=m}^{l}(\mu_{\alpha}^{2}-\mu_{\alpha+1}^{2})=\sum_{\alpha=m}^{l}2\alpha
\end{equation*} 
\noindent
tuhs, following equation can be obtained
\begin{equation}
\mu_{m}^{2}=(l+m)(l-m+1)
\end{equation}
\noindent
Because $l-(-l)$=2 so $l$ can be natural numbers 0,1,2,... or odd multiplicand of $\frac{1}{2}$ that is $\frac{1}{2}$, $\frac{3}{2}$, $\frac{5}{2}$, ... . For Zeeman effect case, $l$ is arbitrary positive natural numbers.

One can see that for every $l$ value, there is (2$l$+1) type of $m$ value associated with (2$l$+1) orthonormal eigenvector of $L_{z}$.

This result shows that every reducible representation, $D$, of a three-dimensional rotation group can be explained by the sum of irreducible representations $D^{(-l)}\oplus D^{(-l+1)}\oplus ... \oplus D^{(l)}$.

Symmetry group of Hamilton operator ${\mathcal H}$, where
\begin{equation}
{\mathcal H}={\mathcal H}_{0}-\frac{e\hbar B_{z}}{2mc}L_{z}
\end{equation} 
\noindent
which is represented by $U(1)\otimes\{E,I\}$ is a commutative group so that all irreducible representation of this group are one-dimensional representation. Representation of group $U(1)$ is $e^{im\theta}$, $m$=0, $\pm$1, $\pm$2, ..., where $\theta$ express rotation angle respect to $z$ axis. It is clear that irreducible representation of $U(1)\otimes\{E,I\}$ will be given by $\pm$$e^{im\theta}$.

$\it{Character}$ of irreducible representation is given by
\begin{equation}
\label{eq58}
{\large \chi}=\sum_{m=-l}^{l}e^{im\theta}=e^{-il\theta}+...+e^{il\theta}
\end{equation}  
\noindent
Similar with the case of finite group, the character shows hyperfine splitting of degeneracy energy level into (2$l$+1) energy levels. This result, which is given by equation (\ref{eq58}), can be basically concluded from the previous results which were given from equation (\ref{eq51}) to equation (\ref{eq54}).   

\section[CONCLUSION]{\label{sec:level5}CONCLUSION}

All symmetry group which are used to explain normal Zeeman effect are compact group. Representation of compact group and representation of finite group have similar properties. All irreducible representation of compact group are equivalent with unitary representation. Likewise, character of irreducible representation will give dimensions of this irreducible representation.

Representation space of compact group should be a Hilbert space which is necessary to define orthonormal vectors. This type of vector space is called $\textit{separable Hilbert space}$ which was extensively used in Section III. 

Finally, symmetry group, which is used to explain normal Zeeman effect as shown in this paper, can be developed to construct symmetry group explaining anomalous Zeeman effect.

For further studies, we would like to consider the implementations of group theory in Darboux transformations\cite{trisetyarso2009application} which have been shown in the cases of graphene \cite{trisetyarso2012dirac} and cavity quantum electrodynamics \cite{trisetyarso2010correlation}  \cite{trisetyarso2011erratum}.  The scheme can be related to the problem of perturbation theory as well \cite{trisetyarso2013perturbation}. This may contribute into the problem of measurement-based quantum computation  \cite{trisetyarso2009measurement} \cite{trisetyarso2009resources} \cite{trisetyarso2011theoretical}.


%\bibliography{New_QIC}

\end{document}